\documentstyle[aps]{revtex}

\begin{document}
\draft
\title{
	Antiprotonic studies of nuclear neutron haloes
      }
\author{
	 S. Wycech\thanks{Internet address: WYCECH@FUW.EDU.PL},
	 J. Skalski\thanks{Internet address: JSKALSKI@FUW.EDU.PL}, and
	 R. Smola\'nczuk\thanks{Internet address: SMOLAN@FUW.EDU.PL}
       }
\address{
	  So{\l}tan Institute for Nuclear Studies,
	    Ho\.za 69, PL-00-681, Warsaw, Poland
	}
\author{
	 J. Dobaczewski\thanks{Internet address: DOBACZEW@FUW.EDU.PL}
       }
\address{
	  Institute of Theoretical Physics, Warsaw University,
		Ho\.za 69, PL-00-681, Warsaw, Poland
	}
\author{
	 J.R. Rook
       }
\address{
       Particle and Nuclear Physics Laboratory, University of Oxford
       }
\maketitle
\begin{abstract}

Nuclear capture of antiprotons from atomic states is studied. Partial widths
for single nucleon capture events leading to cold residual nuclei are
calculated. Recent CERN experiments that compare the neutron and proton
captures are analysed. Nuclear density distributions at extreme nuclear
surface are calculated and tested against the experimental results.
\end{abstract}

\narrowtext

\section{ Introduction}

Recent  CERN  experiments  with  antiprotonic  atoms \cite{LUB94}
have renewed interest in the question of the comparison of the proton
and neutron density distribution at the
surface of large nuclei. It has been known for years that hadronic,
in particular kaonic atoms, provide a way to study the extreme tail of
the nuclear density distribution, its isospin structure and nuclear
correlations there \cite{WIL59}. Two methods have been used
\cite{BUG73,DAV67,BUR67,WIE74,ROB77,KOH86,ROH86}, each of which gives
information in the region roughly 2.0 fm beyond the half density radius.

   1) An observation of the X-ray cascade in hadronic atoms and extraction
of the atomic level shifts and widths. These level widths, related
to the nuclear absorption of hadrons may determine high moments
of the nuclear density distributions. Information obtained in this way is
rather limited, however. For a given atom only one, and in some special
cases two level widths and one level shift, can be measured. The latter is
in general  difficult to interpret and provides a check on the
hadron-nucleon interaction models.

   2) Studies of the nuclear absorption products in particular detection
of the emitted mesons. In this way one can discriminate captures on protons
from captures on neutrons. In principle, more nuclear information is offered
but it is also more difficult to interpret. The reason is that the
initial states of capture are not known directly and the final mesons may
exchange charge or become absorbed.

   The new experiments of Ref. \cite{LUB94}, involving antiprotons, have
two very strong advantages. Firstly they differentiate reasonably clearly
between the  $\bar{p}p$ annihilations and $\bar{p}n$ annihilations, thus
ensuring that the neutron and proton distributions can be separately
estimated. Secondly, they involve antiproton absorbtion very much further
from the nuclear half density radius than the earlier experiments mentioned
above. We shall show that the absorption
occurs around a region 3.0 fm beyond the half density radius.

The idea of the experiments of Ref. \cite{LUB94} is to detect "cold" nuclei,
following the absorption of antiprotons in antiprotonic atoms, by
radiochemical methods. By "cold" here we mean nuclear states of very low
energy, less than the neutron emission threshold. Characterising a nucleus
of $N$ neutrons and $Z$ protons by $(N,Z)$  the reactions are \\

$\bar{p}+ (N,Z) \longrightarrow (N,Z-1) + mesons \hbox{\hspace{2cm}} ({\rm
A})$\\

$\bar{p}+ (N,Z) \longrightarrow (N-1,Z) + mesons \hbox{\hspace{2cm}} ({\rm B})$
\\

Thus $({\rm A})$ involves predominantly interactions of the $\bar{p}$ with
protons and $({\rm B})$ similarly with neutrons.
It is this feature which is the first
advantage of this method mentioned above.

The extreme surface nature of the process arises firstly from the high
orbital angular momentum of the $\bar{p}$, but this is well known and is
exploited in the earlier experiments
\cite{BUG73,DAV67,BUR67,WIE74,ROB77,KOH86,ROH86}.
However the new feature, giving increased surface emphasis arises from
the mesons in $({\rm A})$ and $({\rm B})$. On average there are 4-5 mesons
emitted
and to leave a "cold" nucleus they must all avoid hitting the nucleus.
This can only be achieved in the far surface.

The basic $N\bar{N}$ interactions required for this study are limited
by several phenomenological parameters - range of the $N\bar{N}$ annihilation,
absorptive parts of the scattering amplitudes, pion production
multiplicities, pion momentum distributions. These are taken from
other experiments while  effects of uncertainties must be quantified.
Once the final and initial states are under control one can interpret
the $(N-1)/(Z-1)$ ratios in terms of "neutron haloes" or "neutron skins "
and attribute quantitative meaning to these terms. Qualitatively, the CERN
experiments \cite{LUB94} discover large neutron excess at the nuclear surface
of several heavy nuclei. It complements similar findings in the subcoulomb
neutron pickup reactions \cite{KOR71}.

One purpose of this paper is to provide a description of the nuclear capture
of the atomic antiprotons. Of main interest is that fraction of single
nucleon captures which leaves cold nuclei in the final states. This is the
scenario of the recent experiment \cite{LUB94}. Difficulties in
the way are twofold. First, the $N\bar{N}$ annihilation is a complicated
process with many degrees of freedom involved. Second, the initial atomic
state of the antiproton and the final states of the residual nuclei are not
certain. Fortunately, the first difficulty is moderated by a large energy
release in the $\bar{p}$ absorption. Hence, a closure over nuclear final
states and high energy approximations for the annihilation mesons may
be applied and these yield classical formulas for the absorption rates.
The latter are expressed by integrals of nuclear
densities weighted by a probability to find the antiproton inside nucleus and
a probability to find the final nucleus left undestroyed by the annihilation
products. The last two probabilities are calculated on the basis of the
antiproton and pion optical potentials. Their dependence on the initial
atomic state, final state interactions and parameters of the $\bar{p}$
and pionic nuclear optical potentials are studied in Section II.

Nuclear models of various complication: Fermi gas, shell model, Hartree-Fock
and Hartree-Fock-Bogolyubow methods are used in Section III to find
the nuclear density distributions. Comparison with the experiment is done
and dependence on nucleon binding energies, Coulomb barriers, shell structure
and correlations is studied. The advantages and limitations of the experimental
method are indicated.

\section{ Nuclear absorption of atomic antiprotons}

     Antiprotons bound into atomic orbits cascade down to be ultimately
absorbed by the nucleus. The latter happens at the extreme nuclear surface
and the absorption probability is significant even at distances as large as
twice the nuclear
radius, \cite{BUG73,KOH86,ROH86}. Two effects create such a situation. First,
the free path of antiprotons in nuclear matter is less than 1 fm and second,
the atomic cascade tends to populate states of high  centrifugal barrier $l$.
The peripherality  of capture  allows for standard low density
simplifications: quasi-free scattering and single-particle picture of the
nucleus. Also, it facilitates the description of the final mesons, a vital
question
for understanding the absorption experiments. On the other hand the
disadvantage and difficulty inherent in the surface studies is related to
its sensitivity to range effects.

This section presents a description of the antiproton absorption mechanism.
First, a simple phenomenological picture based on the optical potential
model is presented. Next, two special questions: final state interactions
and range effects are discussed, again in a phenomenological way. A more
detailed justification of the phenomenological approach, basic assumptions
and limitations are given in two consecutive subsections. These contain
rather technical discussion  which may be omitted by readers more interested
in the nuclear structure results.

The tool to describe the antiprotonic atomic level shifts and widths is
an optical potential $V^{\rm opt}$. The simplest one is assumed usually
\cite{KOH86,ROH86,BAT81} in the form
\begin{equation}
\label{1}
      V^{\rm opt}({\bf R})=\frac{2\pi}{\mu_{N\bar{N}}}t_{N\bar{N}}\rho({\bf R})
\end{equation}
where $\mu_{N\bar{N}}$ is the reduced mass, ${\rho({\bf R})}$ is a nuclear
density at a radius ${\bf R}$ and $t_{N\bar{N}}$ is a complex scattering
length. As there is some finite range in the ${N\bar{N}}$ interaction the
density ${\rho({\bf R})}$  involved in Eq.(\ref{1}) is not the "bare"
nucleon density ${\rho_{0}({\bf R})}$ but a folded one
\begin{equation}
\label{2}
  \rho({\bf R)}=\int d{\bf u}\rho_{0}({\bf R}-{\bf u}) \upsilon({\bf u})
\end{equation}
where $\upsilon$ is a formfactor that represents the $N\bar{N}$ force range.
For the absorptive part of $V^{\rm opt}$ the annihilation
range of 1 fm might be
expected from models of the ${N\bar{N}}$ annihilation but the range in the real
part is more difficult to control.

The length $t_{N\bar{N}}$ in Eq.(\ref{1}) is extracted from antiprotonic
atomic data.
The most precise X-ray measurements have been done for the 3d and
4f states in the oxygen isotopes, \cite{KOH86,ROH86}, and fits to these
give $t_{N\bar{N}}$  of about $-1.5-i2.5$ fm, \cite{BAT81,BAT87}. This
value yields a deep and strongly absorptive potential well. At the nuclear
centre Im$V^{\rm opt}$ would be 200 MeV strong and the related
free path length would be
well below 1 fm. However, it should be kept in mind that both the form
and the strength of $V^{\rm opt}$ is tested only in the surface region.
In particular, Im$V^{\rm opt}$ is determined by the atomic level widths,
via
\begin{equation}
\label{3}
  \Gamma=4\frac{\pi}{\mu_{N\bar{N}}}{\rm Im} t_{N\bar{N}}
       \int d{\bf R}\rho({\bf R})\mid\Psi_{\bar{N}}({\bf R})\mid^{2}
\end{equation}
where $\Psi_{\bar{N}}({\bf R})$ is the atomic wave function.
Since $\Psi_{\bar{N}} \approx R^l$ and only high angular momenta $l$ are
available, the absorption strength is peaked at the surface.

The nuclear absorption scenario in $^{58}$Ni is visualised in Fig. \ref{fig1}
where some absorption densities $W=\rho\mid\Psi\mid^{2}R^{2}$ are plotted.
There are two special atomic states singled out in the capture process.
One is the so called "upper" level which usually is the last that can be
detected before the cascading down $\bar{p}$ is absorbed. One can learn
the width of this upper state measuring the intensity loss of the X-ray
transitions. In $^{58}$Ni, and in many other nuclei, the nuclear absorption
is most likely to happen from this level. The next circular state below is
called the "lower" one. Sometimes, one can measure the shape of the
X-ray lines feeding this lower state. Such measurements are possible when
the lower state width is in the range of a few keV and additionally the rate
of radiation from
the upper state competes successfully with the upper state absorption rate.
Chances for antiprotons to reach the "upper" levels have not been measured.
Presumably a significant fraction of the antiprotons is absorbed in states of
much higher principal quantum number $n$. Such a conclusion is reached in
the kaonic atoms where the X-ray transition intensities for the lower atomic
per stopped kaon are known to be about 0.2 to 0.6, \cite{WIE74}.
For the studies of $n/p$ ratios however the value of $n$ is not very important,
one
needs to know mainly the distribution of capture states in terms of the angular
momentum $l$, as at nuclear distances all atomic wave functions of the same
$l$ differ only by irrelevant normalisations. Calculations, \cite{EIS61},
indicate the dominance of $l=l_{\rm upper}$ states in the nuclear capture
process. We discuss the uncertainties in this respect in the last section,
here we calculate
the absorption probabilities assuming full
occupation of the $l=l_{\rm upper}+1$
circular level. The capture probabilities from several circular states are
given  in Table I. The lower
widths are usually larger than upper widths by two orders of magnitude.
That is due to the smaller orbit radii and reduced centrifugal barriers.
However, the absorption density profile is not changed dramatically as may
be seen from the ratio of these densities given in Fig. \ref{fig1}.

The $^{58}$Ni nucleus is our reference case. Amongst the nuclei
tested by the recent $\bar{p}$ capture experiment, \cite{LUB94} see Table II,
it is the simplest to describe.

The localisation of nuclear capture depends on the range of $N\bar{N}$ forces.
One way to find this range is to use ${N\bar{N}}$ potential models, another
perhaps more advantageous, is to fit the atomic and low energy
scattering data. An early choice was to use charge density profiles for
the $\rho$ , \cite{KOH86,ROH86,BAT81}. This is equivalent
to folding a formfactor $\upsilon({\bf u})$ in Eq.(\ref{2}) of 0.8 fm rms
range. More recently, longer ranged
gaussian profile formfactors $\exp(-(r/r_{0})^2)$ have been used,
\cite{BAT87,FRI86}. Typical best fit values are:  $r_{0i} \approx$ 1 fm
(for Im$V$) and $r_{0r} \approx$ 1.5 fm (for Re$V$).
On the other hand, calculations based on the ${N\bar{N}}$ potentials
yield average ranges $r_{0i}$ of 0.75 fm up to 1.45 fm, \cite{GRE82,GRE87},
the difference beeing due to different handling of the off-shell extensions.
An effect of the range is shown in Fig. \ref{fig1}. A longer ${N\bar{N}}$
absorption radius broadens the region of nuclear absorption. The related
effect on the $n/p$ ratio measurement is shown in Fig. \ref{fig2} and
discussed later.

Optical model calculations based on the ${N\bar{N}}$ interaction potentials
\cite{GRE82,SUZ84,GER85,DUM86} indicate that the lengths $t_{N\bar{N}}$ bear
no simple relation to the ${N\bar{N}}$ S-wave scattering lengths which are
smaller and repulsive i.e. with real parts positive. Thus, Re$t_{N\bar{N}}$
is of a complicated and uncertain structure. At the extreme nuclear surface
it reflects a long attractive tail of the pion exchange forces, about
the nuclear radius it may turn to repulsion due to  repulsive scattering
lengths, and is rather uncertain at the nuclear matter densities. On the
other hand, the phenomenological best fit Im$t_{N\bar{N}}$ represents
cumulative effect of the S and P wave absorptive amplitudes, and can be well
understood in terms of the free Im$t_{N\bar{N}}$. The calculated
optical potentials indicate structure more complicated than that given
by formula (\ref{1}), but cannot reproduce the data as accurately as the
latter with the best fit parameters.
In this calculation we use the phenomenological approach.

The level widths discussed so far reflect all modes of the nuclear absorption
of antiprotons. The initial stage, an elementary ${N\bar{N}}$ annihilation,
generates an energy of 2 GeV of which 3/4 is the kinetic energy taken by the
final state mesons. The mesons may excite the residual nucleus via inelastic
scattering and absorption. To calculate the total widths one sums over
the unobserved nuclear excited states. The large energy release and
peripherality allows to use closure over the nuclear states. As a
consequence the effective ${\rm Im} t_{N\bar{N}} $ is close to the absorptive
part of free ${N\bar{N}}$ scattering amplitude.
That is no longer true when the final nuclear states are limited to some
particular states, as is the case of radiochemical methods that detect
"cold" nuclei. The experiments in question, \cite{LUB94}, allow only final
nuclei excited up to the neutron separation threshold. Residual nuclei of
higher excitations would decay by a neutron emission and would not be
detected by the radiochemistry. In the next subsections the spectrum
of allowed excited states
is related to the rearrangement - and to interactions of final state mesons.

Now, to explain our aim we give a simplified result, which will be proven
and refined later. Let ${s}$ denote quantum numbers of the
annihilated nucleon (neutron or proton), the antiproton atomic orbital
$(n,l,j)$ and the final state of the residual nucleus (any nucleus or cold
nucleus). We are going to prove a simple expression for the partial absorption
width $\Gamma_{s}$  corresponding to the "cold" nucleus formation. In the
zero force range limit this reads

\begin{equation}
\label{4}
  \Gamma_{s}=4\frac{\pi}{\mu_{N\bar{N}}}{\rm Im} t^s_{N\bar{N}}
       \int d{\bf R} \mid\Psi_{\bar{N}}({\bf R})\mid^{2}
       \rho_{s}({\bf R}) P_s({\bf R})
\end{equation}
In this expresion a function $P_s$ is introduced to describe formation of the
required final states. It is a product of
two terms $P_s=P_{\rm miss}*P_{\rm dh}$.
The dominant factor in $P_s$ is
$P_{\rm miss}$ - the probability that the mesons born
at point $\bf R$ do not excite the residual nucleus, (missing probability of
Ref. \cite{LUB94}). The other final
state factor - $P_{\rm dh}$ is related to the
final nucleus rearrangement and happens to be less significant.
Examples of these functions are given in Fig. \ref{fig2}. The integrand
in Eq.(\ref{4}), including the $R^2$ factor, is the absorption density for
those processes that lead to the cold $(A-1)$ residual nuclei. It is shown
as $A$ in Fig. \ref{fig2}, in comparison to the full absorption density
$W$ given in Fig. \ref{fig1} it is shifted to the periphery by almost 1 fm.

Now, we derive Eq.(\ref{4}), calculate
$P_{\rm miss}$ , $P_s$ and study the range effects.

\subsection{ Nuclear ${N\bar{N}}$ annihilation and final state interactions}

The aim of this section is to calculate the rate of nuclear ${\bar{N}}$
annihilations that lead to cold final nuclei. This is done in several steps:

1) An amplitude for the ${N\bar{N}}$ annihilation into mesons
$t_{N\bar{N} \rightarrow M}$ is assumed and introduced into the nuclear
transition amplitude in the impulse approximation.

2) The emission probabilities are calculated and summed over the mesonic
and nuclear final states. For an isolated ${N\bar{N}}$ annihilation this
procedure would produce the absorptive cross section and, via the unitarity
condition, the absorptive amplitude Im$t_{N\bar{N}}$. For nuclear captures
leading to cold nuclei we limit the summation over final states to the
states of elastic meson-nucleus scattering. This limited summation generates
the Im$t_{N\bar{N}}$ again, but now it is folded over nuclear final state
interaction factors.

3) To simplify our considerations, effects related to finite range of the
reaction: propagation of the final mesonic resonances, recoil effects,
nonlocalities due to external fields and the size of mesonic source are
discussed at the end of this section.

Assume, that an antiproton in an $n$-th atomic state annihilates on a
nucleon in a single particle state $\alpha$ into ${k}$ mesons with momenta
{\bf $p_{i}$} , $i=1$ to $k$. In the impulse approximation, the transition
amplitude for this process, is
\begin{equation}
\label{5}
      A_{n,\alpha}=\int \Psi^n_{\bar{N}}({\bf x}) \varphi^{\alpha}_N({\bf y})
	 t_{N\bar{N} \rightarrow M}({\bf x},{\bf y},{\bf \xi})
	 \prod_{i}\bar{\varphi}_M({\bf p_{i}},{\bf \xi_{i}},{\beta})
\end{equation}
where $\Psi_{N}({\bf x})$ is the atomic, $\varphi_{N}({\bf y})$
the nuclear, and $\varphi_{M}$ the mesonic wave function. The latter
describes scattering states and corresponds to the ingoing boundary
condition. The final state of the nucleus is not specified, in the spirit
of the impulse approximation it is the initial nucleus that is left with
a hole in the single particle state ${\alpha}$. Additional nuclear
excitations follow nonelastic interactions of the mesons. States generated
in this way are denoted by index $\beta$ in the mesonic wave functions.

The transition  amplitude
$t_{N\bar{N} \rightarrow M}({\bf x},{\bf y},{\bf \xi})$
is not known in detail. What one needs for atomic studies is, basically,
the elastic Im$t^s_{N\bar{N}}$ extended off-shell. The momentum extension is
related to the range dependence expressed in terms of the ${N\bar{N}}$
relative coordinate {\bf x}-{\bf y}. As already discussed this is fairly well
known, contrary to the range dependence in the mesonic coordinates
${\bf \xi}$. The kinematic conditions are special since both the $N$ and
${\bar{N}}$ are bound and the pair energies fall below the $N\bar{N}$
threshold. Even at these energies, the nuclear momenta reach 1$-$2 fm$^{-1}$
and the scattering matrix should include at least S and P waves with all
possible spin states. The relevant partial cross sections or partial
absorptive amplitudes Im$t_{N\bar{N}}$ could be calculated from potential
models of the $N\bar{N}$ scattering. Such a procedure is adopted
in some optical potential calculations, \cite{GRE82,SUZ84,GER85,DUM86,KRO84},
even-though the partial wave analysis of the $N\bar{N}$ scattering does not
exist. Unfortunately, the problem discussed here is more involved and
uncertainties are larger. We aim, rather, at a semi-phenomenological
"effective ${\rm Im} t_{N\bar{N}}$" as used in the phenomenological optical
potential.

To calculate the absorption widths, the amplitudes (\ref{5}) are mod-squared,
summed over the final pionic channels and integrated over the phase space.
One has to sum also over the final nuclear states. In this way one arrives
at an expression for the partial absorption widths $\Gamma_{s}$
\begin{equation}
\label{6}
  \Gamma_{s}=4\frac{\pi}{\mu_{N\bar{N}}} \int \Psi_{\bar{N}}({\bf x})
		 \varphi^{\alpha}_{N}({\bf y})
		  I^{s}({\bf x},{\bf y},{\bf x'},{\bf y'})
		 \bar{\Psi}_{\bar{N}}({\bf x'})
		 \bar{\varphi}^{\alpha}_{N}({\bf y'})
\end{equation}
where
\begin{equation}
\label{7}
 I^{s} = \sum_{\beta} \sum_{k}\int dL \int d{\bf \xi} \int d{\bf \xi'}
 t^{s}_{N\bar{N} \rightarrow M}({\bf x},{\bf y},{\bf \xi})
\varphi_M({\bf p},{\bf \xi},{\beta})\bar{\varphi}_M({\bf p},{\bf \xi'},{\beta})
 \bar{t}^{s}_{N\bar{N} \rightarrow M}({\bf \xi'},{\bf x'},{\bf y'})
\end{equation}
Here, the integration $dL$ means pionic Lorentz invariant phase space
restricted by the energy conservation and $k$ is the pion multiplicity.
For an isolated $N\bar{N}$ system
at or below the threshold, $I^{s}$ in Eq.(\ref{7}) is related by the
unitarity to absorptive part of the elastic $N\bar{N}$ scattering amplitude

\begin{equation}
\label{8}
 I^{s} =  {\rm Im} t^{s}_{N\bar{N}}({\bf x}-{\bf y},{\bf x'}-{\bf y'})
	  \delta({\bf R}-{\bf R'})
\end{equation}
where ${\bf R}=({\bf x}+{\bf y})/2$ is the $N\bar{N}$ CM coordinate.

For an annihilation inside a nucleus this free space unitarity
relation is no longer true. In the external nuclear field the propagation
of intermediate particles in Eq.(\ref{7}) changes. If one is interested
in the total annihilation rates the summation extends over the mesonic as
well as over all the nuclear states. In such a case the closure approximation
applies to nuclear states. It is justified by the large energy release,
peripherality of capture and short annihilation range. Detailed calculations,
\cite{GRE82,SUZ84,GER85,DUM86,KRO84}, have been done for ${\bar{N}}$
optical potentials with models that simulate mesonic channels by
complex ${N\bar{N}}$ potentials. These indicate difficulties of
this question at nuclear matter densities. The effects of the external nuclear
field become noticeable already at $\rho=0.1\rho(0)$ but reflect mainly on
the ${\rm Re} V^{\rm opt}$. The antiprotonic atom physics discussed here is
located safely below this density limit. This qualitative picture justifies
the success of optical potential calculations that can relate the overall
antiproton absorption rates to the free values of ${\rm Im} t_{N\bar{N}}$.
On the other hand, to describe the experimental results corresponding to
the measurements of Ref. \cite{LUB94} one
limits the sum in Eq.(\ref{7}) to cold final nuclei. First we discuss the
chance that the mesons created in the annihilation would leave the residual
nucleus in such states of low excitations. Another excitation mode,
the rearangement is discussed afterwards.

The spectrum of mesons consists essentially of pions correlated in a sizable
fraction into $\rho$ and $\omega$ resonances. These heavy mesons are very
broad and after some \mbox{1 fm} propagation range turn to the pions.
Their multiplicities range from 2 to 8 with an average 4$-$5 and the average
momentum is as large as 2 fm$^{-1}$. Nuclear interactions of these
pions may be absorptive, inelastic or elastic. Those involving a pion
absorption occur on two or more nucleons and produce $(A-2)$ or lighter
nuclei. The inelastic processes end up with excited nuclei. These may be
the $(A-1)$ nuclei of interest as the dominant mechanisms involve single
nucleon excitation modes: $\Delta$ and higher resonances.
For medium and heavy nuclei, and pionic energies around $\Delta$
the inelastic cross sections reach 0.5 b \cite{ASH80}. The main strengths
are located much higher than neutron emission thresholds, however.
Cross sections for excitations of states below these limits are typicaly
1-10 mb, \cite{SHI80}. On the other hand the
elastic cross sections are very large and reach 1 b, \cite{ASH80}.
Hence, the rate for production of cold nuclei is given essentially by
the elasticaly scattered waves. This allows an optical potential description.
In addition, in the bulk of phase space the pions are fast enough
to allow an eikonal description. Following this the wave function for each
pion is taken in the form
\begin{equation}
\label{9}
      \bar{\varphi}_M^{(-)}({\bf p}{\bf \xi}) =
      \exp(i{\bf p}{\bf \xi}-iS({\bf p},{\bf \xi}))
\end{equation}
with $S$ calculated in terms of the pion-nucleus optical potential
\begin{equation}
\label{10}
	 S({\bf p},{\bf \xi}) =
  \int_0^{\infty} ds (\sqrt{(p^2-U^{\rm opt}({\bf \xi}+\frac{{\bf p}}{p}s)}-p)
\end{equation}
The function $S$ is calculated in a quasi-classical way by integrating
the local
momentum over the stright line trajectory. Due to nuclear excitations
and pion absorptions this wave is damped with a rate described by Im$S$.
The latter is generated by absorptive part of the pionic optical potential
Im$U^{\rm opt}$. This damping follows the whole path of a pion but the main
effect comes from regions of large nuclear densities and not the region
around the birth place ${\bf \xi}$. We assume that all functions
$S({\bf p},{\bf \xi})$ are related to the central point of annihilation
${\bf R}$ which is the ${N\bar{N}}$ CM coordinate. Effects of the source size
are discussed later, jointly with consequences of the heavy mesons
propagation range. The assumption on the emission of mesons from the
central point substantially simplifies our calculations. With the mesonic
wave functions (\ref{9}) which enter Eq.(\ref{7}), the total momentum
of mesons ${\bf P}$ separates to a plane wave form. Some additional
dependence on ${\bf P}$ is still there but as we show later it is rather
weak. One consequence is that the ${N\bar{N}}$  CM "conservation"
$\delta({\bf R}-{\bf R'})$ which arises in the free unitarity relation
(\ref{8}) is also a good approximation in the
nuclear case (\ref{7}). Now the final state pion interaction factors that
enter Eq.(\ref{7}) may be collected into a function
\begin{equation}
\label{11}
  P^{k}({\bf R})=< \prod\mid\exp(-S({\bf p_i},{\bf R}))\mid^2 >
\end{equation}
which is a product of the eikonal factors within each multiplicity $k$.
It has to be averaged over the multiplicities and the pionic phase space
weighted by some unknown momentum dependence generated by
$t_{N\bar{N} \rightarrow M}$. The expectation is that the momentum
dependence of $P^{k}$ is weak as compared to the momentum dependence
of $t_{N\bar{N} \rightarrow M}$ since the former is determined by the
nuclear size and the latter by the size of annihilation region.
Thus
one may expected the unitarity condition to hold approximately provided
the averages of $P^{k}$ in Eq.(\ref{7}) are factored out. This averaging
is now performed and in this way one arrives at
  \begin{equation}
 \label{12}
  I^{s} \approx {\rm Im} t^{s}_{N\bar{N}}({\bf x}-{\bf y},{\bf x'}-{\bf y'})
  \delta({\bf R}-{\bf R'}) P_{\rm miss}({\bf R})
  \end{equation}
where the "missing probability" is given by
\begin{equation}
\label{13}
      P_{\rm miss}({\bf R})= \sum_{k} w_{k} \int dL  f(p_{i})
       \prod\mid\exp(-S({\bf p_i},{\bf R}))\mid^2 /
       \sum_{k} w_{k} \int dL  f(p_{i})
\end{equation}
The integration extends over the restricted Lorentz phase space  weighted
by an experimental multiplicity distribution $w_{k}$ , for $k$ from 2
to 8,\cite{CUG89,ARM80}. In order to check the assumptions, some factors
$f(p_{i})$ have been introduced into Eq.(\ref{13}), while pure phase space
and constant $t$ matrices correspond to $f=1$.
This probability density selects those pionic interactions that do not excite
the residual nucleus.

Calculations of $P_{\rm miss}$ are performed in a Monte Carlo procedure.
The optical potential for pions must cover a wide momentum range from the
threshold up to 0.9 GeV but the phase space favours a region just above the
$\Delta$ resonance. This potential is related to the pion nucleon forward
scattering amplitudes and in this way to the pion nucleon cross sections.
That method is well established around the $\Delta$ , \cite{LIU78}.
Here, this procedure is extended to cover also higher $N^{*}_{11}$,
$N^{*}_{13}$ resonances which are described by Breit-Wigner
amplitudes. The two nucleon absorption mode is taken in a phenomenological
form \cite{GIN78}. Performing these calculations one finds that:
high energy expansion of the square root in Eq.(\ref{10}) is satisfactory,
higher resonances cannot be neglected and the black sphere limit is a good
approximation in dense regions. In particular $P_{\rm miss}({\bf R})$ is
changed by less than 10 percent with an inclusion of the
two nucleon absorption. The latter beeing of the $\rho^{2}$ profile
operates in the region where the black sphere limit is well fulfiled.
An example of calculated $P_{\rm miss}({\bf R})$, given in Fig. \ref{fig2},
is close to a pure geometrical estimate that relates it to the solid angle
of the nucleus viewed from the point $\bf R$, \cite{CUG89}. Nevertheless,
the gray zone at the nuclear surface make the effective radius of
an equivalent black sphere  difficult to predict off-hand.

The $P_{\rm miss}({\bf R})$ calculated in this way is at best
semi-quantitative,
but the proximity of strong absorption limit makes the result
fairly independent on the details of the annihilation. One question is
that the phase space alone does not reproduce the experimental momentum
distribution of a single pion, \cite{CLO82}. To remove the discrepancy some
factors $f(p_{i})$, which generate the correct distribution, have been
introduced into average (\ref{13}). New, corrected in this way
$P_{\rm miss}({\bf R})$ is shown in Fig. \ref{fig2}, it is seen to
differ only slightly
from the pure phase space result. As the corrective procedure is uncertain
and the change is below experimental uncertainties we follow the pure phase
space averaging.

In the surface region of interest, the missing probabilities
$P_{\rm miss}$ rise linearly with the distance $R$. That is a fortunate
result, it makes $P_{\rm miss}$ rather
insensitive to the size and structure of the annihilation region located in
a small sphere around ${\bf R}$. The same applies to effects of the $\rho$
and $\omega$ mesons. These may propagate some distance to a point ${\bf R'}$
and decay into pions there. Those events are approximately confined
to within a sphere centered at ${\bf R}$ of a radius = velocity * lifetime
$\approx$ 1 fm. Again, the linearity of $P_{\rm miss}({\bf R'})$ makes an
averaged pionic missing probability equal to the $P_{\rm miss}({\bf R})$.

Let us turn now to other corrections.
The annihilation happens at the nuclear surface and is confined to a region
of a small diameter. Nevertheless, effects of the nucleus should be
considered. These are: the $N\bar{N}$ centre of mass motion, external field,
and Pauli principle in the intermediate states. At distant surface
most of these have been found small in the optical model calculations.
Now we discuss briefly the implications in the $A-1$ reactions.
The first, elementary effect due to the presence of nucleus is the
${N\bar{N}}$ centre of mass motion with respect to the residual nucleus.
It is given by the recoil energy i.e. by a Fourier transform of
$t_{N\bar{N}}(E-P^2/4M)$ over the the ${N\bar{N}}$ CM momentum $P$.
It is known in
the kaonic atoms \cite{DAV67,BUR67} that a narrow (30 MeV or less)
resonance close to the threshold would induce a propagation range as large
as 1 fm and affect strongly the peripherality of nuclear capture.
Extensive experimental efforts gave no clear evidence for narrow
resonances in the $N\bar{N}$ system close to the threshold. Energy independent
transition matrices $t_{N\bar{N} \rightarrow M}$ are assumed here. Hence
the annihilation is fast and the $N\bar{N}$ CM may be fixed in the absorption
process, thus a factor $\delta({\bf R}-{\bf R'})$ is justified in
Eq.(\ref{12}). On the other hand, an effect of the CM motion arises due
to dependence of the pionic functions $S({\bf p_{i}},{\bf R})$ on the total
momentum of mesons ${\bf P}$. It induces some nonlocality in the
${\bf R}-{\bf R'}$ but the effect enters in a second order of a small
quantity $P$/(pion momentum). Numerical studies described earlier
indicate a nonlocality of a 0.1 fm range, negligible in comparison
to the 1 fm range effects involved in the relative coordinate
${\bf x}-{\bf y}$.

\subsection{ Range effects }

     In the limit of zero range $N\bar{N}$ interactions formula (\ref{6})
for the partial absorption width may be expressed in terms of nuclear
densities. For finite-ranged interactions the single particle wave
functions involved may be reduced only to mixed densities. However, for
simplicity and historic reasons one wants to have an approximate
expression in terms of the true densities. At the nuclear surface this
can be done with a good precision, at least for the  absorption
rates summed over all nucleon states. The standard relation, \cite{CAM78},
that allows it is:
\begin{equation}
\label{14}
  \sum_{\alpha} \bar{\varphi}^{\alpha}_{i}({\bf y'})
				  \varphi^{\alpha}_{i}({\bf y})
  \approx \rho_{i}({\bf Y})j_{0}(k_{\rm F}({\bf Y})\mid {\bf y}-{\bf y'} \mid)
\end{equation}
where ${\bf Y}=({\bf y}+{\bf y'})/2$ , $k_{\rm F}$ is an effective local Fermi
momentum which may be calculated in a shell model and $i=p,n$.
In a similar way we express the angular averaged atomic wave
functions $\bar{\Psi} \Psi$  by
\begin{equation}
\label{15}
\frac{1}{2l+1}\sum_{m}\bar{\Psi}^n_{\bar{N}}({\bf x'})\Psi^n_{\bar{N}}({\bf x})
\approx
\mid\Psi^n_{\bar{N}}({\bf X})\mid^{2}\{D^{n}({\bf X}-{\bf X'})
			 +O({\bf X}/{nB},{\bf X'}/{nB})\}
\end{equation}
It is an expansion in the inverse Bohr radius of the orbit $1/nB$ and
calculation of function $W$ is strightforward.

Finally, to handle the $N\bar{N}$ ranges we assume a separable approximation
for the scattering matrix in Eq.(\ref{12})
$t_{N\bar{N}} = \upsilon ({\bf x}-{\bf y},r_{0}\sqrt{2})
t^0_{N\bar{N}} \upsilon({\bf x'}-{\bf y'},r_{0} \sqrt{2})$ and
gaussian \mbox{form-factors} $\upsilon$ with some range parameters
$r_{0}$ . This allows for a simple transformation to relative coordinates
and reduces Eq.(\ref{6}) into a folded density expression, \cite{GRE82}.
For the capture rates one has now:
\begin{equation}
\label{16}
\Gamma_{s}= 4\frac{\pi}{\mu_{N\bar{N}}}{\rm Im} t^0_{N\bar{N}}
	    \int\mid\Psi_{\bar{N}}({\bf Y})\mid^{2}
	      \upsilon({\bf Y}-{\bf X},r_{0}) f_{X}
	      \rho_{i}({\bf X}) P^{s}_{\rm miss}(\frac{{\bf X}+{\bf Y}}{2})
\end{equation}
where $f_X$ is a factor that collects together a large piece of finite
range effects
\begin{equation}
\label{17}
  f_{X}=\int d{\bf u}\upsilon({\bf u},2r_{0})j_{0}(k_{\rm F}({\bf X}){\bf u})
						      D({\bf u})d{\bf u}
\end{equation}
With a normalised form-factor $\upsilon$ the limit of zero range force is
$f_X=1$. For a typical $N\bar{N}$ absorption range $r_{0}=1$ fm values
of $f_{X} \sim 0.5-0.6$ are obtained. These are almost constant in the
nuclear surface region of interest and may be taken out of the integral.
As our analysis involves only ratios of the widths the actual values of
$f_X$ are unimportant.

     The summation  in Eq.(\ref{14}) selects capture events that lead to
single hole nuclear states. These do not correspond to the experimental
conditions that require cold final nuclei or, more precisely, nuclei
either in the ground states or in states  excited below the neutron
emission threshold $T_n$ . To account for it we limit the sum over the
initial nucleon states in Eq.(\ref{10}) to those that leave the final
nucleus with excitation energies less than $T_{n}+2$ MeV where the 2 MeV
is allowed for neutron kinetic energies. At the surface almost
all single particle states of sizable overlap with the atomic
antiprotons contribute to the sum. The experimental cut-off that eliminates
deeply bound nucleons becomes a (small) correction, which we call
a "deep hole" factor. Defined as the ratio of limited sum of single nucleon
densities to the total sum
\mbox{$P_{\rm dh}(\bf X)=\sum_{\alpha}^{ltd} \varphi_{\alpha}^2/ \sum_{\alpha}
\varphi_{\alpha}^2$ } and implemented into the partial width formula produces
\begin{equation}
\label{18}
  \Gamma_{s}(cold)=4\frac{\pi}{\mu_{N\bar{N}}}{\rm Im} t^0_{N\bar{N}}f_{X}\\
       \int\mid\Psi_{\bar{N}}({\bf Y})\mid^{2}
       \upsilon({\bf Y}-{\bf X},r_{0})
  \rho_{i}({\bf X})P_{\rm dh}({\bf X})P_{\rm miss}(\frac{{\bf X}+{\bf Y}}{2})
\end{equation}

This is our result and now we turn to practical, model calculations of the
basic ingredients in this equation.

\section{ Calculations of nuclear densities}

For the first test we use an  asymptotic  density  (AD)  model.  It
follows, essentially, the Bethe-Siemens approach \cite{BET68} but  it  also
incorporates larger phenomenological input i.e. charge density distribution,
neutron and proton separation energies and difference of the  rms  radii
of proton and neutron densities. At central densities  a  Fermi  gas  of
independent protons and neutrons  is  assumed.  The  Fermi  momenta  are
determined by the densities and the Fermi energies are fixed by the separation
energies. This gives the depth of the potential well which  in  the  surface
region is extrapolated by the Woods-Saxon form. The  densities  are
given by the exponential damping of the nucleon wave  functions  due  to
the potential barriers. For protons  a  Coulomb  barrier  is  added  and
potential parameters (half-density radius $c$ and the surface thickness
$t$) are fitted to reproduce the experimental charge
density down to 5 percent of the central density. For neutrons the  same
$t$ is used but $c$ is chosen to obtain the rms radius equal (or larger
by 0.05-0.10 fm in the heaviest nuclei, \cite{SAT79}) to the proton density
rms radius. This model is expected to generate average level densities,
it misses shell
effects and correlations.

As a second method to determine neutron and proton densities we have
used a selfconsistent Hartree-Fock theory with the effective two-body
Skyrme-type interaction.  Since our aim in using HF
method was rather unusual, i.e. to find nucleon densities at the
extreme tails of the nuclear matter distribution (at distances of 8$-$15
fm from the center), a few remarks about its practical
implementation seem to be in order here.

The necessary practical condition is the use of a HF code not
restricting in any way the asymptotic form of s.p. wave functions. This
condition excludes e.g. all codes using the harmonic oscillator basis.
In the present work we have applied the code solving HF equations on the
spatial mesh, in which all fields and densities are expressed in the
coordinate representation.

The most severe restriction of the presented results is the imposed
spherical symmetry.  It allows enormous
simplification of solution, in particular the HF equation takes the
form of a differential equation in the radial variable for
each pair of the conserved s.p. quantum numbers $l$ and $j$. We used 100 mesh
points in the radial coordinate, in a box of the size of 25 fm and put as
a boundary condition the values of the wave functions at the far end of
the box equal to zero.

     The density matrix is obtained by summing contributions from the
lowest s.p. orbits. If necessary, the
contribution of the last orbit is calculated in the filling approximation;
i.e. an appropriate occupation probability, smaller than one, is
associated with this orbit.

The asymptotic form of the radial s.p. wave function
for large $r$ corresponding to the s.p. energy eigenvalue $\epsilon_{nlj}$ is
\begin{equation}
\label{19}
     {\cal R}_{nlj}(r) = w_l(\frac{1}{r}) \exp(-\kappa r)/r ,
\end{equation}
where $\kappa=\sqrt{\frac{2m}{\hbar^2}|\epsilon_{nlj}|}$ and $w_l$ is a
polynomial with the dominant term being a constant, so that the
contribution to the asymptotic density is proportional to
$\exp(-2 \kappa r)/r^2$. For proton orbitals there is an additional
exponential factor, coming from the Coulomb barrier and modifying $\kappa_p$,
which is very important for distances in question, i.e., between 7 and 15 fm.
Although at very large distances from the center the neutron to proton
density ratio is proportional to $\exp(2 (\kappa_p-\kappa_n) r)$,
where $\kappa_n$ and $\kappa_p$ are directly related to
neutron and proton separation energies, respectively, at distances
near the ${\bar p}$ absorption peak usually a few neutron and proton
orbitals contribute significantly to the density and a more detailed
analysis is necessary to evaluate the latter.

S.p. binding energies (Fermi levels) important in determination of
nuclear density tails are not reproduced exactly with existing
effective Skyrme forces. In addition, calculated spherically symmetric
densities for deformed nuclei lack quadrupole correlations which
{\it a priori} may for their own sake distort positions of the Fermi levels.
Therefore, calculated densities, especially for deformed nuclei, must
be treated as approximate, and possible sources of error must be kept in
mind. In particular, the true densities of deformed nuclei may have
longer tails since their elongated form has to be smeared over Euler angles
in order to obtain spherically symmetric density in the LAB frame.

In spite of the approximate character of our nuclear density calculations
we used also HFB theory \cite{Bog,RS} in order to assess modifications
introduced to HF results by the residual pairing interaction.
In the present case it must be
distinguished from the very often used HF+BCS method, in which the
pairing correlations are included using the BCS prescription to
self-consistent orbitals. In the latter case, the partial occupation of
orbitals above the Fermi level leads to a non-zero, though usually
minuscule, occupation of orbitals of positive energy (in continuum).
Since such orbitals are not localized this implies that they dominate
nuclar density at large distances which is a completely unphysical
effect.

The applied HFB code also uses the coordinate representation and
the HFB equations are solved on a spatial mesh. The proper analysis of
the asymptotic properties of two-component quasiparticle wave functions
shows \cite{DFT} that the HFB ground state wave function, even
containing pairing correlations, is always localized if bound.
The asymptotic form of the occupied negative-energy quasiparticle states is as
in the Eq.(\ref{19}), with $\kappa$ defined in terms of the
sum $E_{nlj}-\lambda$, where $E_{nlj}$ is the quasiparticle
energy and $\lambda$ is the Fermi energy.

As the effective force we use the ten-parameter Skyrme SkP interaction
described in Ref. \cite{DFT}.  It has a virtue that the pairing matrix
elements are determined by the force itself, contrary to other
Skyrme-type interactions which define only the particle-hole channel.
In the paired HFB ground-state the pairing gap is state dependent.
As a simple pairing gap parameter one can use
the pairing potential average over the occupied states.

The "deep hole" corrective factor in the HFB
method is calculated using
the additional condition in the form
$\epsilon_n < \lambda$, where $\epsilon_n$ are the expectation values of the
self-consistent mean-field matrix (not q.p. energies) and $\lambda$ is
the HFB Fermi energy.

The last method we used for determination of nuclear densities was chosen
to assure correct separation energies.  A single particle (SP) spherical
well, including the central and spin-orbit
potentials for  neutrons and protons was assumed. Proper order of s.p. levels
is guaranteed by the form of the potential.
Potential parameters were adjusted slightly to obtain
the experimental separation energies, charge rms radii and, if known, the
neutron rms radii.

\section{ Results}

In this section we discuss the partial antiproton absorption widths:
$\Gamma_{n}(A-1)$ and $\Gamma_{p}(A-1)$ for captures on a neutron and
proton, respectively, which produce cold $(A-1)$ final nuclei.
Sum of the two is denoted by
$\Gamma(A-1)$. Experiments determine those partial widths relative to
the total absorption width $\Gamma^{\rm tot}$. The data collected in Table II
consist of two such ratios: $\sigma_{A-1}=\Gamma(A-1)/\Gamma^{\rm tot}$
and $\sigma_{np}=\Gamma_{n} (A-1)/\Gamma_{p}(A-1)$. The first one
$\sigma_{A-1}$ is a test for a description of the antiproton absorption.
In particular it checks the weakest points: understanding of the final state
interactions and knowledge of the initial atomic states of capture.
If quantitative understanding is achieved one can claim control over
the region of nuclear surface where the neutron halo is measured. The
halo itself is seen via the $\sigma_{np}$ ratio.

\subsection{ The $\sigma_{A-1}$ ratios }

A typical antiproton absorption scenario is visualized in Fig. \ref{fig1}
and Fig. \ref{fig2}, which contain some ingredients of formula (\ref{18})
for the capture widths. The results, given in Table I, show $\sigma_{A-1}$
calculated for some circular atomic orbits that are most likely to be the
states of nuclear capture.
The shapes of $(A-1)$ capture densities are determined by the angular momentum
$l$, strong nuclear absorption and $P_{\rm miss}$
and thus are rather independent
on the normalisation of the atomic wave functions i.e. on  $n$. Thus, these
ratios are typical to all $n$ states.
With the angular momentum $l$ increased by one unit, in particular
from the "lower" to the "upper" and higher $l$ states, the $\sigma_{A-1}$
increases by about 20\%. Thus, the experimental data exclude sizable
fraction of high $l$ captures but seem less restrictive on the
states with $l$ lower than $l_{\rm lower}$ ,
where the calculated $\sigma_{A-1}$
stabilises. On the other hand, cascade calculations done in kaonic, hyperonic
and antiprotonic atoms, \cite{WIE74,EIS61}, show the nuclear capture from
the latter very low $l$ states to be unlikely. Also, these calculations
indicate an accumulation of the capture probability on
two or at most three values of $l$.
This result is consistent with our capture probabilities, given in Table I,
and calculated under the extreme assumption that
the $l=l_{\rm upper}+1$ circular level is fully occupied at some stage of the
atomic cascade. On the other hand, calculations of Ref. \cite{ROB77}
allow a broader distribution with a 20 percent share of the
$l_{\rm upper}+1$ and higher $l$ states and a 10 percent share of
$l_{\rm lower}-1$ values.
If that is the real situation our results for $\sigma_{A-1}$ and
$\sigma_{np}$ would rise typically by a factor of 1.05. This is the likely
uncertainty of the calculations in Tables I, II due to poor knowledge
of the capture orbits. One hopes to clarify some of these points by
experimental measurements of the cascade intensities and absolute cascade
intensities in the nuclei of interest, \cite{PRO95}.

The $\sigma_{A-1}$ calculated with the AD and other models are consistent
with most of the experimental data, shown in Table II. This gives some
confidence in the validity of the final and initial state description.
However, there are two outstanding discrepancies: Te and Yb.

{\em Special cases of  $^{130}${\rm Te} and $^{176}${\rm Yb} }

The first case is understood qualitatively, the second presents a point of
specific interest. It is known experimentally \cite{WYC93} that a strong
E2 mixing, i.e. coupling of the atomic and nuclear rotations, occurs in the
upper level of the $^{130}$Te atom. It stimulates absorption from the upper
level as indicated in Table I, and induces an effect of alignments of the
nuclear and atomic spins in the states admixed to the upper level. Thus the
orbital antiproton stays closer to the elongated part of the nucleus, as
compared to states of equally averaged orientations.
Thus, the final pions have a better chance to miss the nucleus.
Calculations yield some 20\%
enhancement of the total absorption widths due to this effect \cite{WYC93}.
One expects similar enhancement of the $\sigma_{A-1}$ rate, it is also
likely that the $n/p$ ratio is higher at the poles of this nucleus.
The strong E2 mixing happens also in $^{176}$Yb for high $n=14$ atomic
orbits, however. It is not clear as yet what are the consequences for
the atomic cascade process between the $n=14$ state and nuclear absorption.
It is also not clear what correlation of the atomic motion and nuclear
orientation is induced by this effect. Future experimental and theoretical
studies \cite{PRO95} will help to elucidate this point.

Antiproton absorption in the heaviest elements, $^{232}$Th and $^{238}$U,
is accompanied by a nuclear fission of the final $(A-1)$ nuclei which, in
principle, may affect the $\sigma_{A-1}$ rate. However, in such nuclei
the radiative rates dominate the fission rates for excitations less than
the neutron emission threshold, \cite{WHE58}. In the even-odd nuclei of
interest this domination is even stronger. Thus, the fission channel
is expected to change the $\sigma_{A-1}$ only a little, and this is
apparently borne out by the data in Table II.

\subsection{ The $\sigma_{np}$ ratios }

The partial absorption widths are proportional to effective absorptive
amplitudes ${\rm Im} t_{N\bar{N}}$ for the $\bar{p}n$ and $\bar{p}p$ pairs.
These are not well known, although some average values follow from the
optical potential phenomenology. The number required for neutron halo
studies
is a ratio $R_{np}= {\rm Im} t(\bar{p}n)/{\rm Im} t(\bar{p}p)$
which may be taken from
other experiments. One value $R_{np}=0.63 $ has been obtained by Bugg,
\cite{BUG73}, from measurements of charged pions emitted in the $\bar{p}$
absorption in Carbon. Difficulties arise since it
includes effects of final state mesonic interactions and the inherent
uncertainties of the charge exchange reactions. This value of $R_{np}$
generates mild disagreement with the data of Ref. \cite{LUB94} for all the
nuclei and all nuclear models used here.  The results given in Table I
should be compared to the experimental data in Table II, similar
discrepancies are generated by other models. A different result
$R_{np}=0.81(3)$ follows from the stopped antiproton absorption in
deuterium, \cite{BIZ74}. This value is free from the pionic effects,
but the deutron kinematic conditions, in particular the binding
energy, differ from those met at the nuclear surface.
Another value of $R_{np}$ obtained
in $^{4}$He is smaller and energy dependent. At rest, a number
$R_{np}=0.48(10)$ has been deduced from rather involved analyses of the
final state mesonic interactions in the three nucleon systems \cite{BAL}.

In this work we fix $R_{np}$ from a best fit to our simplest
nucleus which is $^{58}$Ni. Our nuclear models yield similar results
in this case and the fitted $R_{np}$ is very close to the value obtained
from the deuteron. We shall use the latter in our analysis.

The results for $\sigma_{np}$ are collected in Table II. It is clear that
the crudest model of asymptotic density strongly overestimates the
$\rho_n/\rho_p $
ratios at large distances. This property has been known already from
the neutron pickup studies, \cite{KOR71}. By the same effect, the AD model
produces too large $\sigma_{A-1}$ in the heaviest nuclei Th and U.
The physics behind it is quite transparent:

1) It is vital to have correct separation energies but these are not the
whole story.

2) The Coulomb barriers enhance anomalously the $\rho_n/\rho_p$
ratio at large distances.
That has to be off-set by shell effects (angular momentum barriers)
and correlations.

3) Proper setting of the neutron skin defined in terms of mean squared radii
$R_{ms}(neutrons)-R_{ms}(protons)$ does not determine the "neutron halo".
The latter are understood here in terms of $\sigma_{np}$ i.e. ratios of
high moments of density distributions.

What are the moments involved in the haloes measured by the radiochemical
experiment?
For zero range interactions and $P_{\rm miss}=1 $
these are the "Barret moments"
i.e. $2l$-th moments due to centrifugal barriers corrected for the atomic
wave functions. The $P_{\rm miss}$ and $P_{\rm dh}$ increase
the order of the moments
approximately by two units. On the other hand, the annihilation range
effects, i.e., the folding, introduces moments smaller by two, four and more
units. The joint effect is best estimated by the dominant $2l$ density
moment involved. As we see from Table I these are very high moments
ranging from 10 in Ni to 18 in U.

The HF method with the SkP force gives roughly correct neutron separation
energies for (nearly) spherical systems $^{58}$Ni, $^{96}$Zr,
$^{144}$Sm, underbinds the last neutron in $^{96}$Ru by 1.3 MeV but
generally underestimates proton separation energies, e.g. by 1.1 MeV in
$^{144}$Sm and by 3 MeV in $^{58}$Ni. This statement is qualitatively true
also for the rest of deformed nuclei, with the one exception of $^{154}$Sm
where the last neutron is underbound by 2.9 MeV. As we have checked in
a separate calculation the Skyrme force SIII does not improve description
of Fermi energies in the studied nuclei.

Comparison between the data and HF results shows clear disagreement for Yb
and $^{144}$Sm and less pronounced one for Te, Th, and U nuclei. On the
basis of comparison of calculated and experimental separation energies
one can expect qualitatively corrections to the calculated $\sigma_{A-1}$
and $\sigma_{np}$ ratios (remembering that like errors in proton and neutron
Fermi energies tend to compensate each other for the $\sigma_{np}$). In all
presented cases except $^{144}$Sm and $^{176}$Yb they go in the right
direction.

In order to better understand the asymptotic HF densities one can look
closer at how many orbitals contribute to it at large distances
and to test by means of the formula (\ref{14}) the sensitivity of
the $\rho_n/\rho_p$ ratio to shifts in s.p. energies.
In $^{58}$Ni, at $r=7$ fm, two orbitals give about 35\% of the neutron density
each, 3 others give about 10\% each. For protons, there is one orbital
contributing 56\% (the highest one) and three other contributing
about 12\% each. At $r=15$ fm, the last occupied orbitals contribute
89\% and 91\% to the neutron and proton densities, respectively. Of course,
the heavier the nucleus, the farther the asymptotic region and the
more orbitals contribute to one-body densities in the range 8-15 fm.
In $^{144}$Sm, at $r=7$ fm, 4 states contribute more than 10\% each to
the neutron density (the highest percentage being 24\%), four others
contribute more than 5\% each. Even at 15 fm, there are still 4 neutron
orbitals contributing significantly to $\rho_n$ (32\%, 24\%, 20\% and 16\%),
the reason being that their s.p. energies differ not more than 4 MeV. In
$^{238}$U, at 15 fm, one must account for 4 neutron orbitals while only two
proton orbitals contribute 79\% and 15\% of the $\rho_p$.
Clearly, it is not only a distance $r$ but also the (sub)shell structure
which decides how many orbitals contribute.

In order to estimate the effect of s.p. energies on the density ratio
$\rho_n/\rho_p$ we have used the following procedure. Contributions to
the density at $r=8$ fm were used as a data and the propagation to $r=15$ fm
was performed using Eq.(\ref{19}), with account for the Coulomb barrier.
The resulting densities for $^{96}$Zr are then larger than
the exact HF densities
roughly by the factor 7/4 and 9/7 for neutrons and protons, respectively.
(The $\rho_n/\rho_p$ ratio is then 161 instead of 115.) This error comes from
the influence of the polynomial $w_l$ in Eq.(\ref{19}) but we concentrate on
the effect of changing s.p. energies on the so calculated $\rho_n/\rho_p$
at $r=15$ fm. The decrease of the neutron energies by 1 MeV changes
this ratio to 94, the decrease of the proton energies by 1 MeV rises
the ratio to 234 and the simultaneous decrease of both proton and
neutron energies by 1 MeV slightly decreases the ratio to 137.
This gives some feeling as to the sensitivity of the $\rho_n/\rho_p$
to the neutron and proton binding energies.

{\em Special case of  $^{58}${\rm Ni} }

This is our reference nucleus. The input data is certain and all model
calculations produce consistent results. We use $R_{ms}(n)=3.734$ and
$R_{ms}(p)=3.710$ , \cite{SAT79}, which in the SP approach produce
$\sigma_{A-1}=1.05$ and $\sigma_{np}=0.88$ . That is close to the HF
and HFB results from Table II.
The latter two methods are not perfect, HF underbinds protons by
0.5 and neutrons by 0.3 MeV. These errors are of no significance, however,
since the absorption is spread over four neutron and four proton s.p.
states.

{\em Special case of  $^{144}${\rm Sm} }

This nucleus displays a proton halo. Qualitatively, it might be due to
the closed neutron shell. The separation energy of neutrons is large
(10.6 MeV) with respect to a small (6.2 MeV) one for the protons.
However, this proton halo effect is not reproduced in our
calculations. Boths
results for  $\sigma_{np}$ given in Table II and plots of the
$neutron/proton$ density ratios given in Fig. \ref{fig3} indicate
a neutron excess at the
surface. The separation energies in our nuclear models are either
fitted to the experimental (AD or SP) or well reproduced for neutrons.
For protons the HF model underestimates the experimental
value by 0.9 MeV
and that should even enhance the proton tail over the real one.
This case indicates again that it is not only the separation energy that
matters for the nuclear tail. The experimental result in $^{144}$Sm
is not understood and opens the case for more exotic speculations.

One obvious effect of inclusion of pair correlations is
a change in the $\rho_n/\rho_p$ ratio in the tail of the density
following from the change in the Fermi energy. At smaller distances
however, this ratio may change in the opposite direction if other levels
than the last one contribute to the density.
It turns out that in all studied cases the change in the $\rho_n/\rho_p$
due to pairing is small up to 14 fm. A much more pronounced pairing effect
on both ratios $\sigma_{np}$ and $\sigma_{A-1}$ comes from the $P_{\rm dh}$
factor which e.g. in $^{232}$Th changes from 0.52 (no pairing) to 0.69
(with pairing) at the total absorption peak at $r=8$ fm.
This change more than balances a decrease in $\rho_n/\rho_p$ ratio providing
for larger values of $\sigma_{A-1}=0.109$ and $\sigma_{np}=4.65$
(see Table II). The same $P_{\rm dh}$ factor
is responsible for an increase of both ratios in $^{238}$U, to 0.100 and
4.31, respectively, while the smaller $P_{\rm dh}$ leads to smaller
$\sigma_{A-1}$ and $\sigma_{np}$ in $^{96}$Zr (0.117 and 2.40, respectively).
Pairing changes also $\sigma_{A-1}$ for $^{154}$Sm to 0.106 and
$\sigma_{np}$ for $^{154}$Sm and $^{176}$Yb to 2.96 and 3.50,
respectively. Changes due to pairing are nearly none for other nuclei.

\section{ Conclusions}

The radiochemical method which detects the products of nuclear capture of
antiprotons is a valuable source of information on the relative
$neutron/proton$ density distribution on the extreme tail of nuclear surface.
The main features which we want to stress are:

1) The nuclear regions tested are more peripheral than those studied
by the X ray measurements in hadronic atoms. One measures essentially
the $2l$ moments of the density distributions where $l$ is the angular
momentum of "upper levels". These moments are as high as 18 in
the heaviest nuclei.

2) There are special cases of alignment of nuclear and atomic angular
momenta formed by the E2 mixing which display higher $n/p$ ratios and higher
rates of cold single nucleon captures. These may test the composition of
the pole regions in deformed nuclei. Further studies are recommended.

3) The uncertainty in the initial atomic state of capture is kept under
a fair control by the $\sigma_{A-1}$ rates. However, additional X-ray
experiments would be helpful to clarify this question.

4) Strong neutron haloes are observed in heavy deformed nuclei. These
are not determined by the binding energies and Coulomb barriers alone.
The shell effects (angular momentum) are also important. Few (two or three)
of the highest nucleon orbitals contribute most to the cold capture rates.

5) An interesting case of a proton halo is found in $^{144}$Sm.
It is not understood in terms of single particle models, and may signal
strong nuclear correlations in the surface region.

6) Apart from the  $^{144}$Sm case the nuclear models reproduce the
qualitative features of the observed haloes.

\vspace{5mm}
We thank J. Jastrz\c{e}bski for his constant interest and
collaboration and P.E. Hodgson for his advice and for arranging the
collaboration that was kindly supported by The Royal Society and KBN.

\newpage
\widetext
TABLE I. Atomic results. Column 2 contains the principal q.number $n$
and the angular momentum $l$. For the remaining columns:
c.p. is the nuclear capture probability calculated under the assumption
that the circular atomic state $n=n_{\rm upper}+1$ is fully occupied,
$\sigma_{A-1}$= $\Gamma^{A-1}/\Gamma^{\rm tot}$is the branching ratio for
the cold capture  and $\sigma_{np}$= $\Gamma_{n}/\Gamma_{p}$ is the ratio of
captures on neutrons to protons.  The AD model and $R_{np}=0.63$ were used.
\\[5mm]

\begin{tabular}{cccccc}
\hline
\hline
ELEMENT & n & l & c.p. & $\sigma_{A-1}$ & $\sigma_{np}$  \\
\hline
		&   &   &      &       &        \\
  $^{58}$Ni     &  4& 3 & 0.   & 0.095 & 0.69   \\
		&  5& 4 & 0.16 & 0.097 & 0.69   \\
		&  6& 5 & 0.83 & 0.110 & 0.70   \\
		&  7& 6 & 0.01 & 0.150 & 0.71   \\
		&  8& 7 & 0.   & 0.220 & 0.71   \\
  $^{96}$Zr     &  6& 5 & 0.24 & 0.106 & 4.67   \\
		&  7& 6 & 0.72 & 0.128 & 5.30   \\
 $^{130}$Te     &  7& 6 & 0.05 & 0.096 & 1.77   \\
		&  8& 7 & 0.93 & 0.122 & 2.00   \\
 $^{144}$Sm     &  7& 6 & 0.01 & 0.075 & 1.39   \\
		&  8& 7 & 0.75 & 0.085 & 1.46   \\
 $^{154}$Sm     &  7& 6 & 0.01 & 0.087 & 3.65   \\
		&  8& 7 & 0.75 & 0.099 & 3.98   \\
 $^{176}$Yb     &  8& 7 & 0.23 & 0.097 & 3.34   \\
		&  9& 8 & 0.75 & 0.124 & 4.07   \\
 $^{232}$Th     &  7& 6 & 0.   & 0.073 & 3.94   \\
		&  8& 7 & 0.   & 0.091 & 4.64   \\
		&  9& 8 & 0.31 & 0.098 & 5.00   \\
		& 10& 9 & 0.69 & 0.127 & 6.20   \\
 $^{238}$U      &  9& 8 & 0.29 & 0.106 & 6.55   \\
		& 10& 9 & 0.71 & 0.138 & 8.24   \\
		&   &   &      &       &        \\
\hline
\hline
\end{tabular}
\vspace{1cm}

\widetext
TABLE II. Comparison of nuclear models. Experimental and
calculated results for $\sigma_{A-1}$ and $\sigma_{np}$
are given. Calculations for atomic orbitals weighted as
in Table I are done with $R_{np}=0.82$ .\\[5mm]

\begin{tabular}{ccccccccc}
\hline
\hline
& \multicolumn{2}{c}{Exp.\cite{LUB94}} & \multicolumn{2}{c}{AD}
& \multicolumn{2}{c}{HF} & \multicolumn{2}{c}{HFB}\\
		&          &        &      &   & &  & & \\
 & $\sigma_{A-1}$ & $\sigma_{np}$ &
   $\sigma_{A-1}$ & $\sigma_{np}$ &
   $\sigma_{A-1}$ & $\sigma_{np}$ &
   $\sigma_{A-1}$ & $\sigma_{np}$ \\
\hline
            &           &        &      &      &      &       &      &       \\
  $^{58}$Ni & 0.098(8)  & 0.9(1) & 0.11 & 0.90 &0.110  & 0.785 &0.110 & 0.781
\\
  $^{96}$Zr & 0.161(22) & 2.6(3) & 0.12 & 4.9  &0.125  & 2.54  &0.117 & 2.40
\\
  $^{96}$Ru & 0.113(17) & 0.8(3) & 0.10 & 1.7  &0.099  & 0.944 &0.099 & 0.955
\\
 $^{130}$Te & 0.184(36) & 4.1(1) & 0.12 & 2.6  &0.124  & 3.14  &0.123 & 3.22
\\
 $^{144}$Sm & 0.117(20) & $<$ .4 & 0.09 & 1.9  &0.094  & 1.38  &0.092 & 1.36
\\
 $^{154}$Sm & 0.121(20) & 2.0(3) & 0.10 & 5.1  &0.110  & 3.34  &0.106 & 2.96
\\
 $^{176}$Yb & 0.241(40) & 8.10(7)& 0.12 & 4.8  &0.111  & 3.23  &0.109 & 3.50
\\
 $^{232}$Th & 0.095(14) & 5.4(8) & 0.12 & 7.6  &0.087  & 3.80  &0.109 & 4.65
\\
 $^{238}$U  & 0.114(9)  & 6.0(8) & 0.13 & 10   &0.092  & 4.09  &0.100 & 4.31
\\
            &           &        &      &      &      &       &      &       \\
\hline
\hline
\end{tabular}
\newpage

\begin{figure}
\caption[F1]{
The total antiproton absorption densities from the "upper" $n=6$ , $l=5$
orbit in $^{58}$Ni: $W_l$ for the ${N\bar{N}}$ annihilation range $r_0=1$ fm
and $W_s$ for the range $r_0=0.75$ fm. The dot-dashed line is a ratio of
two $W_{l}$ for subsequent circular $n=5$ and $n=6$ atomic states.
$\rho_{0}$ is a "bare" neutron density. Normalisations are arbitrary.}
\label{fig1}
\end{figure}

\begin{figure}
\caption[F2]{
The $(A-1)$ "cold" antiproton absorption density on a neutron from $n=6$
circular orbit in $^{58}$Ni. $A_l$ given by the integrand of Eq.(\ref{4})
for the ${N\bar{N}}$ annihilation range $r_0=1$ fm and $A_s$ for the range
$r_0=0.75$ fm. $\rho_{0}$ is a "bare" neutron density. Normalisations are
arbitrary. Missing probabilities (left scale): $P_{\rm miss}$ continuous
is due to phase space alone, $P_{\rm miss}$ dash-dotted is calculated
with corrections for the experimental pion momentum distribution.
The flat dashed curve is $P_{\rm dh}$ from the HFB model.}
\label{fig2}
\end{figure}

\begin{figure}
\caption[F3]{
The ratios of neutron  and proton densities calculated with several
nuclear models in the $^{144}$Sm nucleus. Dashed line-HF,
dotted-SP, dot dashed-AD, continuous line-HFB. The $(A-1)$ "cold"
absorption density peaks at 8.6 fm marked with a dot while its bulk
is located between 7.2 and 10.2 fm. The experimental
$\sigma_{np} < 0.4$ .}
\label{fig3}
\end{figure}

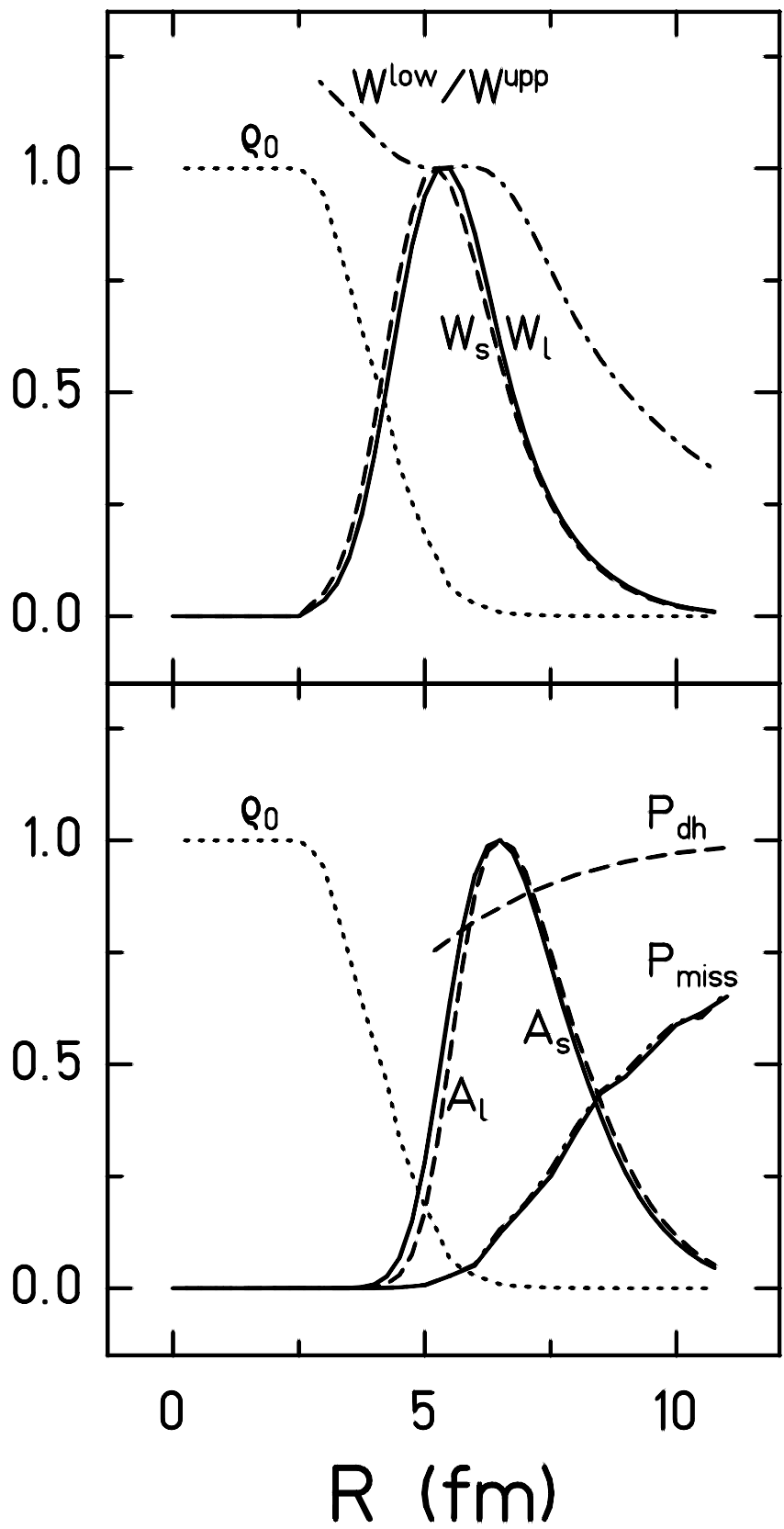
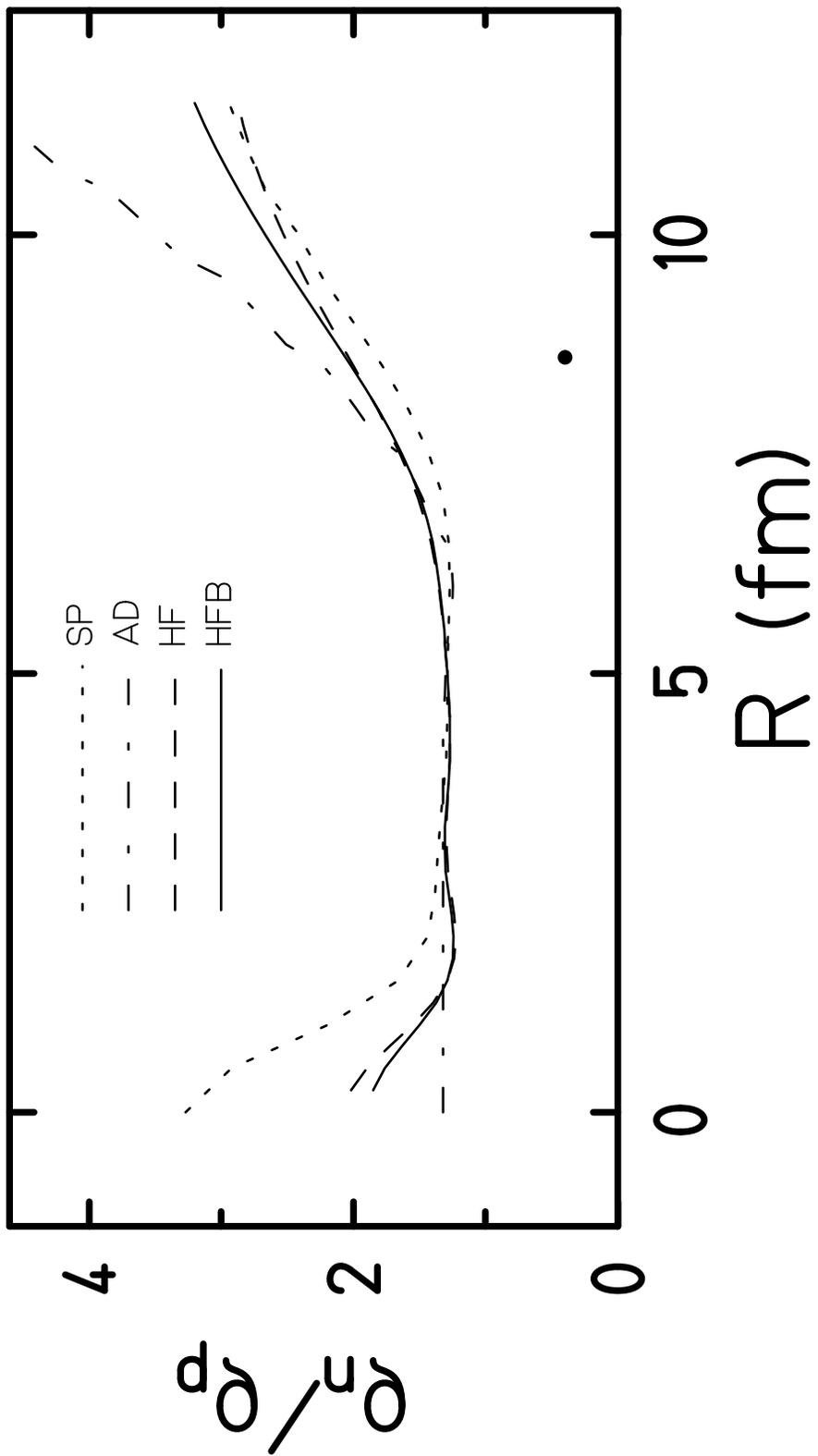
\end{document}